# Molecular Dynamics Simulation of Macromolecules Using Graphics Processing Unit

Ji Xu[1,3], Ying Ren[1], Wei Ge[1*], Xiang Yu[2,3], Xiaozhen Yang[2], Jinghai Li[1]

[1]State Key Laboratory of Multiphase Complex Systems, Institute of Process Engineering (IPE),
Chinese Academy of Sciences, P.O. Box 353, Beijing 100190

[2] Beijing National Laboratory for Molecular Sciences,
Joint Laboratory of Polymer Science and Materials,
Institute of Chemistry, Chinese Academy of Sciences, Beijing 100190

[3]Graduate University of the Chinese Academy of Sciences, Beijing 100039, China;

* Corresponding author, email: wge@home.ipe.ac.cn

**Abstract** Molecular dynamics (MD) simulation is a powerful computational tool to study the behavior of macromolecular systems. But many simulations of this field are limited in spatial or temporal scale by the available computational resource. In recent years, graphics processing unit (GPU) provides unprecedented computational power for scientific applications. Many MD algorithms suit with the multithread nature of GPU. In this paper, MD algorithms for macromolecular systems that run entirely on GPU are presented. Compared to the MD simulation with free software GROMACS on a single CPU core, our codes achieve about 10 times speed-up on a single GPU. For validation, we have performed MD simulations of polymer crystallization on GPU, and the results observed perfectly agree with computations on CPU. Therefore, our single GPU codes have already provided an inexpensive alternative for macromolecular simulations on traditional CPU clusters and they can also be used as a basis to develop parallel GPU programs to further speedup the computations.

## 1 Introduction

Macromolecular system has been one of the most active areas in polymer science and molecular dynamics (MD) simulation has proven to be a powerful computational tool to complement experimental results. However, the computationally-intensive nature of MD algorithms and the limited computational horsepower available today make it difficult for current simulations to reach large spatio-temporal scales as in macromolecular experiments, though several MD packages, such as GROMACS[1-4], NAMD[5], LAMMPS[6], can already run efficiently on distributed memory computer clusters. Fortunately, Graphics processing unit (GPU), originally designed for computationally-intensive, highly-parallel graphic operations, now becomes programmable for general-purpose computations with the advent of convenient software development environments, such as the Compute Unified Device Architecture (CUDA)[7] from NVIDIA. With CUDA, GPU can serve as an accelerator to CPU, executing a very large number of threads in parallel. Some MD algorithms, with atom as the smallest particle, are data-parallel and can be mapped to GPU conveniently with one thread dealing with one atom. Stone et al.[8] accelerated non-bonded force calculation in NAMD codes with a spatial bin method which doesn't require the CPU to build neighbor lists. Anderson et al.[9] developed a general purposed MD program fully implemented on GPU except the part of bins updating and they treated the pair wise interactions of both Lennard-Jones interactions and bonded interactions. These techniques have shed some light on the simulation of macromolecular systems[10].

In this article, we implement MD simulations of macromolecular systems on a single GPU to reach a larger spatio-temporal scale so as to investigate the dynamics of polymer entanglement and crystallization. We have found that the time consumed in memory coping process of bins updating increases with the size of the simulated systems. So in our codes, all the MD processes including the bins updating part are put into the GPU, and angle potentials, dihedral potentials are also included to keep macromolecular conformations.

**2 Model**

In this paper, the macromolecular systems considered are the polymer molecules of polyethylene. The monomers are treated as two types of united-atoms, representing $CH_2$ (C_32) and $CH_3$ (C_33) respectively. Each polyethylene molecule consists of 150 monomers, resulting in 298 C_32 and 2 C_33 interaction sites. The Dreiding II force field[11] is adopted. The interactions for an arbitrary geometry of a molecule are expressed as a sum of internal forces, which include bonded interactions ($E_{bonded}$) and non-bonded interactions ($E_{nb}$). The former depends on molecular structure and the latter depends on the distance between the atoms, that is,

$$E = E_{nb} + E_{bonded}. \tag{1}$$

$E_{nb}$ is calculated with Lennard-Jones potential:

$$E_{nb}(r_{ij}) = 4\varepsilon_{ij}\left[(\frac{\sigma_{ij}}{r_{ij}})^{12} - (\frac{\sigma_{ij}}{r_{ij}})^6\right], \tag{2}$$

where $\varepsilon_{ij}$ is the depth of the potential well and $\sigma_{ij}$ is the finite distance at which the interparticle potential is zero.

Three types of interactions are considered in $E_{bonded}$, namely, bond stretch ($E_b$, two-body), bond-angle bend ($E_a$, three-body) and dihedral angle torsion ($E_d$, four-body), which are called bond, angle, dihedral respectively for simplicity.

$$E_{bonded} = E_b + E_a + E_d, \tag{3}$$

where $E_b$ is expressed as a harmonic potential between atom *i* and *j*:

$$E_b(r_{ij}) = \frac{1}{2}k_{ij}^b(r_{ij} - b_{ij})^2, \tag{4}$$

$k_{ij}^b$ is the force constant and $b_{ij}$ is the equilibrium distance between i and j.

$E_a$ is represented by a harmonic potential on the angle $\theta_{ijk}$:

$$E_a(\theta_{ijk}) = \frac{1}{2}k_{ij}^a(\theta_{ijk} - \theta_{ijk}^0)^2, \tag{5}$$

$k_{ij}^a$ is the force constant and $\theta_{ijk}^0$ is equilibrium angle of atoms i-j-k.

$E_d$ is calculated as

$$E_d = k_\phi(1 + \cos(n\phi - \phi_s)), \tag{6}$$

where $\phi$, n and $\phi_s$ represent the angle between the i-j-k and j-k-l planes, the periodicity, and the equilibrium angle respectively. Parameters in the formula mentioned above are listed in Table I.

Table I. Parameters of the force field

| | Type | $\sigma$ /nm | $\varepsilon$ /(kJ·mol$^{-1}$) |
|---|---|---|---|
| Atom | C_32 | 14.027 | 0.3624 |
| | C_33 | 15.035 | 0.3699 |
| | Type | $k_{ij}$/(kJ·mol$^{-1}$·1k$^{-2}$) | $b_{ij}$/nm |
| Bond | C_32-C_32 | 292880 | 0.153 |
| | C_32-C_33 | 292880 | 0.153 |
| | Type | $k_{ij}$/(kJ·mol$^{-1}$·1kJ·mo$^{-2}$) | $\theta_{ijk}$/degree |
| Angle | C_32-C_32-C_32 | 418.40 | 109.471 |
| | C_32-C_32-C_33 | 418.40 | 109.471 |
| | C_33-C_32-C_32 | 418.40 | 109.471 |
| Dihedral | Type | $k_\Phi$/(kJ·mol$^{-1}$) | $\Phi_s$/degree | n |
| | C_32-C_32 | 4.184 | 0 | 3 |

### 3 GPU-based algorithms

The GPU-based algorithm is developed at IPE based on their previous work[12]. The general simulation procedure is illustrated in Figure 1. Leap-Frog scheme[13] is adopted to integrate the equations of motion. Polymer system is simulated using NVT ensemble and extended ensemble Nosé-Hoover method[14-16] is used to control the temperature. For several reasons, such as force truncation and integration errors, the translation and rotation around the center of mass would be inevitably generated and should be removed during the simulation.

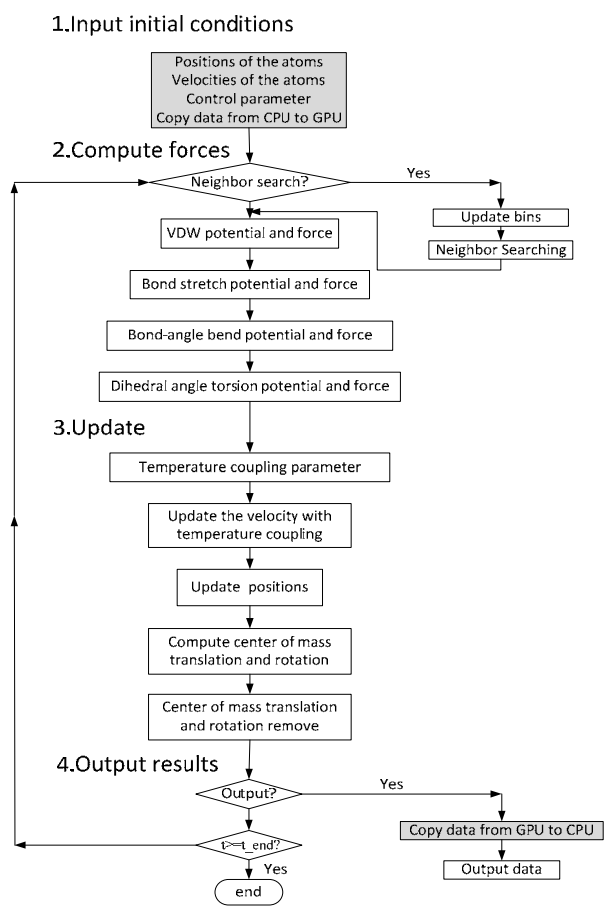

Figure 1. Diagram of the whole MD procedure

Note: Shaded boxes involve data transfer between GPU and CPU.

Here we define 5 variables, as shown in Table II. The atom handled by the global thread is defined as home atom, and atomic information can be loaded from the global memory of GPU by GTIDX.

Table II. Abbreviations of variables

| Variable | Description |
| --- | --- |
| GAIDX | global atom index of the system |
| GTIDX | global thread index of a GPU kernel, GTIDX=GAIDX-1. |
| MAIDX | atom index in one molecule, ranging from 1 to NAOM. |
| MIDX | molecule index |
| NAOM | number of atoms in one molecule |

*3.1 Binning the atoms and neighbor list generation*

As our simulations usually involves a huge number of atoms, grid search is much faster than simple search when generating the neighbor list[17], and is hence adopted. In this approach, the simulated domain is first divided into grids with size of each dimension equal to the cutoff distance of the non-bonded force. All the atoms are then put into the corresponding grids according to their positions. In this way, an atom only needs to search the atoms in its own grid and 26 neighboring grids, totally 27 grids.

To bin the atoms, it is natural and preferred to assign one atom to one thread, since the element of the simulated system is atom, and that of the GPU kernel is thread. However, for earlier NVIDIA GPU, such as Tesla C870, which does not support atomic functions, it is difficult to avoid writing to same global memory simultaneously in this way. On the other hand, if the atoms are binned on CPU first and the binned data are then copied into GPU, coined as CBIN method, significant data transfer between CPU and GPU is required for large systems. In this paper, we present two methods, GBIN_OLD and GBIN_NEW, designed for the earlier and recent generation of GPU respectively.

In GBIN_OLD, binning is still implemented on GPU. To achieve relatively good performance, a method with three steps is proposed, named as GBIN_OLD. The main data structure used is:

```
typedef struct{
    int Nmax;        // the max number of atoms that one bin contains
    int *idx_bins;   // dim: Nr, the grid indices of the atoms
    int Mx,My,Mz;    // the number of the grids
    int *size;       // dim: Mx*My*Mz, the number of particles one bin contains
    int *size_back;  // dim: Mx*My*Mz, the former number of particles one bin contains
    int *idxlist;    // two dimensional array that contains the atom indices
    int *idxlist_back;   // two dimensional array that contains the former atoms indices
}t_bin;
```

And the key processes are illustrated in Figure 2:

Step 1, generate idx_bins array. The bin index of every atom is computed according to its new position. It's easy to compute the grid index of each atom with one thread, which is a totally data-parallel task.

Step 2, delete the atoms in each grid which do not belong to it any more. The element in this kernel is bin rather than atom. The number of thread blocks is the number of bins, but each thread block contains only one thread, making kernel inefficient. Before deleting the atoms out of the bins, the information of former bins must be backed up first, that is, the size and idxlist arrays are copied to size_back and idxlist_back arrays respectively.

Step 3, add new atoms into the bins. Information of the 26 neighbor bins of the former bins stored in idxlist_back array is searched to find which atoms have moved into the central bin, as indicated by bin 13 in gray. The indices of the atoms of the neighbor bins are loaded from the idxlist_back array, and the new bin indices of the atoms are obtained from idx_bins. If the new bin index equals to the index of the central bin, the atom is added. This is the most time consuming step among the three.

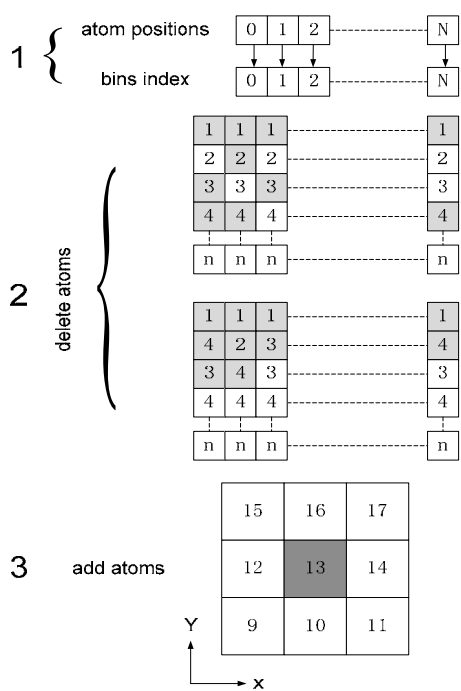

Figure 2. Algorithm of GBIN_OLD

Note: This figure only depicts the slice with the home grid in X-Y plane depicted in step 3.

In comparison, CBIN needs only part of the data for GBIN_OLD, that is,

typedef struct{
    int Nmax;　// the max number of atoms that one bin contains
    int *idx_bins;　// Nr, the grid indices of the atoms
    int Mx,My,Mz; // the number of the grids
    int *size;　// the number of particles one bin contains
    int *idxlist;　// two dimensional array that contains the global atoms indices
}t_bin;

In CBIN, no backup of the size and idxlist arrays is needed, so the usage of global memory is reduced. The positions of the atoms are copied from the global memory of GPU into corresponding CPU memory first, and the bin index of every atom is then computed according to its position, followed by putting the atoms into the idxlist array according to their bin indices. Finally, the binned data are copied back into the global memory of GPU.

Multiple threads can write to the global memory, which starts from the same position, using atomic functions to index the writing position. So GBIN_NEW method is proposed with the support of atomic functions of new generation GPU. In this method, the data needed is the same as CBIN and the procedure is similar too. The core code is as follows:

1. int gtidx = blockIdx.x*blockDim.x + threadIdx.x; int idx_bin = idx_bins[gtidx];
2. int num = atomicAdd(&size[idx_bin], 1);
3. int mem_pos = Nmax*idx_bin + num;
4. idxlist[mem_pos] = gtidx + 1;

In the first line, gtidx is GTIDX. In line 2, atomic operation is used to retrieve the number of atoms already in the bin when this thread visits size[idx_bin] in global memory. Line 3 computes the memory position where the new atom should be stored. The atom index which starts from "1" is stored into the bin in line 4.

A neighbor list of each atom should be made after binning atoms. The cutoff distance ($r_{cut}$) is set to a value large enough so that non-bonded potentials can be fully expressed. A two dimensional array (N*M) in the global memory of GPU is allocated to store the neighbor list. Here N is the number of atoms plus the padded number given by the cudaMallocPitch function, and M is the maximum number of neighbors for each atom. The number of neighbors of every atom is different, especially for heterogeneous systems, which makes it a luxury way to use memory, but the global memory of GPU will be read in a coalesced way[7].

In the neighbor list searching kernel, one thread block corresponds to one gird, and one thread corresponds to one atom in the bin. One thread block deals with itself and the neighboring 26 bins. The positions and atom indices are loaded from global memory to shared memory in advance, and then each thread uses the data in shared memory to search the neighboring atoms.

In Dreiding II force field[11], only the first and second bonded neighbors are excluded from the non-bonded interactions, and only an int4 type data is needed to store the exclusions of one atom. Although reading int4 data from the global memory of GPU can accelerate the speed, not all the force fields only exclude the first and second bonded neighbors. For general purpose, two arrays are used to store the exclusions:

```
typedef struct {
    int *index;
    int *excls;
}t_exclusions;
```

The whole t_exclusions contains all the exclusions of one type of molecule. And index stores the starting position of the exclusions of one atom; excls stores the exclusions of the whole molecule.

Each thread computes MAIDX and MIDX according to GAIDX. So index[MAIDX-1] is the starting position of the exclusions, and (index[MAIDX]-index[MAIDX-1]) is the number of the exclusions of this atom. The exclusions of the atom are loaded into registers from global memory, then these data are added by NAOM*MIDX to get the global exclusions of the atom. Details of these two arrays are given in section 3.3.

*3.2 Non-bonded interactions*

After generating the neighbor list, computation of non-bonded interactions is cast into GPU kernel with one thread dealing with one atom. The codes for non-bonded interactions are outlined as follows:

```
int gtidx = blockIdx.x * blockDim.x + threadIdx.x;
…
float fx=0.0f, fy=0.0f,fz=0.0f;
for(int i=0; i<n_neigh; i++)
{
  …
  float dx = pos.x - neigh_pos .x;
  float dy = pos.y - neigh_pos .y;
  float dz = pos.z - neigh_pos .z;
  …
  float rsq = dx*dx + dy*dy + dz*dz;
  float fforce = 0.0f;

  float r2inv;
  if(rsq >= r_cutsq) r2inv = 0.0f; else r2inv = 1.0f / rsq;
  float r6inv = r2inv*r2inv*r2inv;
  …
  fforce = r2inv*r6inv* (12.0f*ljp2*r6inv - 6.0f*ljp1);
  // accumulate the forces
  fx += dx * fforce;
  fy += dy * fforce;
  fz += dz * fforce;
}
```

First GTIDX (gtidx) is computed according to the executing environment of the GPU kernel. It's also GAIDX-1. Data (position, type, number of neighbors) of the home atom are then loaded into registers according to GTIDX, followed by the inner loop of non-bonded calculation. The parameters of formula (3) and (4) are stored in the GPU constant memory and are loaded through types of the interacting atoms.

*3.3 Bond, angle and dihedral interactions*

In these three types of interactions, bond interactions can be seen as pair-wise. But the number of interaction pairs of bond interactions is fixed while that of non-bonded interactions is changing during the simulation. The angle and dihedral interactions of a certain molecule are also fixed throughout the simulation. For this reason, these three different types of interactions are illustrated together here.

Usually, interaction lists are generated according to the types of molecules at the beginning of the simulation. For the polyethylene system, it involves only one type of molecule. Bond interactions are treated pair-wisely like exclusions, and the data structure is, hence, also similar:

```
typedef struct {
    int Nr_mol; // number of atoms in this type of molecule
    int *index;
    int *bond;
} t_bonds;
```

Here Nr_mol is the number of atoms of different molecules. The array index is the same as t_exclusions, and array "bond" contains indices of atoms which have bond interactions in the molecule. A simple molecule depicted in Figure 3 is taken to describe the details of constructing the "index" and "bond" arrays.

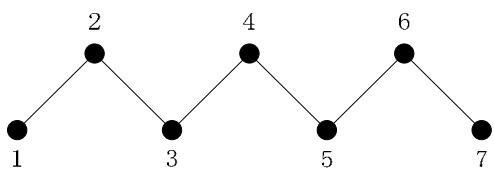

Figure 3. Schematic show of a molecule

Figure 4. Contents of index and bond arrays in t_bond

There're 7 atoms in this molecule, so Nr_mol is set to 7, and "index" has 8 elements with "0" as the first one representing the beginning of the "bond" array. It is also the starting position of the memory storage of the first atom in "bond". The first atom has one bond. The index of the bond atom (2) is stored from "index[0]" (0) in "bonds". The second element of "index" is the value of the element before it plus the number of bonds the previous atom has, so it is 1 and is also the starting position of the memory storage of the bond of the second atom. The second atom has two bonds, 1-2 and 2-3, so the second element of "bond" is set to 1, the third to 3. Likewise, "index" and "bond" are constructed. When t_bonds has been generated, the data are copied to the global memory of GPU during the "Input initial conditions" step in Figure 1.

The bond interactions are computed according to t_bonds. It's convenient to map this work into GPU kernel with one thread dealing with one atom. But there's a significant difference between the algorithms of GPU and CPU. In CPU, all the bond interactions are computed in a for-loop way.

One bond interaction is computed in one loop, and the force can be given to both atoms of the bond. But in GPU, it's difficult to treat bond interactions this way. To make use of the multithread nature of GPU, every thread should compute all the bonds of the home atom, so actually the operations are doubled. The codes used to get all of the bonds of one atom according to its GTIDX are listed below:

```
  int gtidx = blockIdx.x * blockDim.x + threadIdx.x;
  int nr_bonds = 0;
  int my_bonds[MAXBONDS];
  int my_mol_idx = gtidx % Nr_mol;
  int my_mol = int(gtidx / Nr_mol);
  nr_bonds = index[my_mol_idx+1] - index[my_mol_idx];
  int start_bond_pos = index[my_mol_idx];
  for(int i=0;i<nr_bond;i++)
  {
    my_bonds[i] = bond[start_bond_pos+i] + Nr_mol*my_mol;
  }
```

First GTIDX is calculated from the execution environment of the kernel. Then the GAIDX of bonds of the home atom are obtained from the "index" and "bond" data of GTIDX, and stored in my_bonds. Following the bond interactions can be computed.

As to angle and dihedral interactions, the inherent multi-body feature makes them much more complicated than pair-wise interactions. Their data structures are given below:

```
typedef struct{
  int ang_idx;
  int pos;
  int2 ang_atoms;// first and second atom indices of an angle except the home atom
} t_angle_ele;

typedef struct {
  int Nr_mol; // number of atoms in this type of molecule
  int *index;
  t_angle_ele *angles;
} t_angles;
```

The t_angle_ele structure represents one angle interaction, and it is different from the CPU algorithm which only contains indices of the three atoms in an angle. The ang_idx is used to choose the parameters in formula (7). The pos is the position of the home atom in the angle. And ang_atoms are the indices of the other two atoms in the angle. It should be noted that the values in ang_atoms must be in the order from the first to the third atom of an angle. In t_angles, Nr_mol and index have the same meanings with those in t_bonds. According to the definitions above, the contents of t_angles for the molecule in Figure 3 can be depicted as shown in Figure 5.

```
index:           | 0 | 1 | 3 | 6 | 9 |12 |14 |15 |
         ang_idx | 0 | 0 | 1 | 0 | 1 | 2 | 1 | 2 | 3 | 2 | 3 | 4 | 3 | 4 | 4 |
         pos     | 1 | 2 | 1 | 3 | 2 | 1 | 3 | 2 | 1 | 3 | 2 | 1 | 3 | 2 | 3 |
angles:  at1     | 2 | 1 | 3 | 1 | 2 | 4 | 2 | 3 | 5 | 3 | 4 | 6 | 4 | 5 | 5 |
         at2     | 3 | 3 | 4 | 2 | 4 | 5 | 3 | 5 | 6 | 4 | 6 | 7 | 5 | 7 | 6 |
                   ↑   ↑   ↑       ↑           ↑           ↑       ↑
                   1   2   3       4           5           6       7
```

Figure 5. Data structure of t_angles

After the generation of t_angles data, it is copied into the global memory of GPU. And then the angle interactions can be calculated in the same way as bonded interactions. It should also be noted that, in CPU algorithm, the angle interactions are usually computed in for-loop type with one loop dealing with one angle interaction, and then forces are added to the atoms composing the angle. But in GPU algorithm, similar to the calculation of bond interaction, the number of floating point operations is tripled.

Dihedral interactions can be treated in the same way as angle interactions. But the position of the home atom of one interaction is from 1 to 4. And the floating point operations are quadruple compared to the CPU algorithm.

*3.4 Update configurations*

The Leap-Frog scheme[13] together with Nosé-Hoover temperature coupling algorithm is adopted to update the configurations. It is straight forward to map the Leap-Frog scheme into GPU kernel by using one global thread to update the information of one atom. The temperature coupling algorithm, however, needs the kinetic energy of the whole system, which makes it comparatively difficult to be realized in GPU. Here we propose a method to store the kinetic energy of each atom in a temporary array, followed by a summation of the energies using the method in CUDA SDK[18].

**4 Performance evaluations**

The performance of the simulation method discussed above is evaluated on the Mole-8.7 system[19] at IPE using a single HP XW8600 workstation with two Intel® Xeon® E5430 2.66GHz processors, eight 667 MHz 2GB JEDEC RAM, two Nvidia® GeForce GTX 295 GPU cards and one Tesla C870® GPU for the GBIN_OLD . The operating system is CentOS5. The C/C++ compiler is GCC-4.1.2. The GPU driver is NVIDIA–Linux-x86_64-180.29-pkg2.run[20]. The CUDA® SDK used is cuda-sdk-linux-2.10.1215.2015-3233425.run[21], and toolkit package is cudatoolkit_2.1_linux 64_rhel5.2.run[22]. The CPU program used is GROMACS-4.0.5 [4]. As our GPU program, it also uses floating point arithmetic, so the comparison is more meaningful.

To evaluate the performance of each algorithm, 7 polyethylene systems with different volume(*V*) and different numbers of polyethylene molecules (N), as listed in Table III, are simulated and different parts of the algorithms are evaluated respectively.

Table III. Configurations of 7 test systems

| System | I | II | III | IV | V | VI | VII |
|--------|---|----|-----|----|----|----|-----|

| $V$ (nm$^3$) | 10*10*16 | 16*16*24 | 20*20*32 | 24*24*40 | 30*30*40 | 34*34*40 | 40*40*40 |
|---|---|---|---|---|---|---|---|
| $N$ / 1000 | 19.2 | 64.8 | 117.6 | 243.0 | 363.0 | 507.0 | 675.0 |

*4.1 Binning the atoms and generating neighbor list*

The execution time of three methods for binning the atoms and copying binned data between GPU and CPU using different hardware and algorithms is listed in Table IV, where CBIN is further divided into three distinct steps, as previously described in section 3.1. Table IV suggests that, in CBIN, the time of two data copy procedure is longer than that of binning. The GBIN_OLD method takes much longer time than CBIN, but the GBIN_NEW method is more efficient, and the larger the system is, the more time would be saved. Compared to GBIN_OLD, this method not only reduces the use of the global memory of GPU but also speeds up binning atoms.

Table IV. Execution time (μs) of binning and copying binned data between GPU and CPU

| | System | I | II | III | IV | V | VI | VII |
|---|---|---|---|---|---|---|---|---|
| CBIN | position copy | 0.112 | 0.353 | 0.634 | 1.160 | 2.763 | 4.061 | 5.022 |
| | binning | 0.168 | 0.566 | 1.043 | 3.288 | 6.086 | 8.445 | 11.539 |
| | binned data copy | 0.384 | 1.286 | 2.384 | 3.916 | 4.824 | 5.867 | 8.008 |
| | total | 0.664 | 2.205 | 4.061 | 8.814 | 13.673 | 18.373 | 24.569 |
| GBIN_OLD | | 3.934 | 12.328 | 22.067 | 43.584 | 64.178 | 88.749 | 123.468 |
| GBIN_NEW | | 0.484 | 4.853 | 1.636 | 3.347 | 5.096 | 6.938 | 9.3516 |

Table V. Execution time (μs) of neighbor list generation

| System | I | II | III | IV | V | VI | VII |
|---|---|---|---|---|---|---|---|
| GPU | 8.199 | 22.628 | 37.382 | 67.62 | 102.681 | 130.05 | 181.691 |
| CPU1 [4] | 20.870 | 70.445 | 124.246 | 257.868 | 385.575 | 545.796 | 723.741 |
| CPU2 [4] | 45.339 | 159.842 | 277.442 | 563.646 | 857.880 | 1182.425 | 1571.300 |

The GPU times of the neighbor list generation listed in Table V are taken from our own codes, and the CPU times are taken from GROMACS-4.0.5. The major difference between CPU1 and CPU2 is how to set the charge groups. Every ethylene has one charge group in CPU1, while every atom has one charge group in CPU2. Since neighbor searching is based on charge groups in GROMACS-4.0.5, CPU1 has less floating point operations and runs faster than CPU2. Here we suggest comparing GPU with CPU2 since they have the same floating point operations.

*4.2 Computation of non-bonded interactions*

Non-bonded interactions are computed after the generation of the neighbor list. The computation times of the 7 systems of both GPU and CPU programs are given in Table VI.

Table VI. Execution time (μs) of non-bonded interactions

| System | I | II | III | IV | V | VI | VII |
|---|---|---|---|---|---|---|---|
| GPU | 1.316 | 4.088 | 7.278 | 14.761 | 21.839 | 30.305 | 40.250 |
| CPU | 7.637 | 24.544 | 44.715 | 92.989 | 138.969 | 193.915 | 258.405 |

Apparently, the GPU program is about 6 times faster than the CPU program. Compared to some other GPU programs which reach tens or even hundreds times of speedup, our result doesn't seem quite good. This may be ascribed to the following reasons. First, in the kernel of non-bonded interaction, the calculation has only 20 floating point arithmetic operations, but each inner loop reads the global memory of GPU three times and the constant memory twice. Taking into account the huge difference in speed of the floating point arithmetic operations and global memory reading of GPU, the speedup can not be high. Second, the non-bonded interaction is pair-wise and can be computed only once in CPU algorithm. However, in GPU kernel, this feature can not be used easily, so the total floating point arithmetic and memory reading of GPU kernel almost doubled. In ref [9], with about 102 neighbors of one particle on average, the execution time of one step is about 3 ms with 12,000 particles in the system. There're about 45 neighbors of one particle in our 675,000 testing system, taking the faster speed of new generation GPU into account, our code is still fast enough.

*4.3 Computation of bonded interactions*

The running times of the three types of bonded interactions for GPU and CPU kernels are given in Table VII, VIII and IX respectively. It can be found that GPU kernels are much more efficient than corresponding CPU kernels. For CPU kernels, although the numbers of bond, angle and dihedral interactions decrease slightly, the running times increase significantly due to the increase of the interaction complexities, meanwhile the times increase with the size of the system. However, for GPU kernels, the running times do not change much for the three different types of interactions, or for different sizes of the system, indicating that the intensities of computations are very close to each other.

Table VII. Execution time (μs) of bond interaction

| System | I | II | III | IV | V | VI | VII |
|---|---|---|---|---|---|---|---|
| GPU | 0.089 | 0.258 | 0.448 | 0.914 | 1.362 | 1.893 | 2.513 |
| CPU | 0.855 | 2.909 | 5.258 | 11.051 | 16.518 | 23.012 | 30.899 |

Table VIII. Execution time (μs) of angle interaction

| System | I | II | III | IV | V | VI | VII |
|---|---|---|---|---|---|---|---|
| GPU | 0.243 | 0.783 | 1.409 | 2.905 | 4.331 | 6.06 | 8.045 |
| CPU | 2.942 | 9.951 | 18.096 | 37.288 | 56.063 | 78.259 | 103.933 |

Table IX. Execution time (μs) of dihedral interaction

| System | I | II | III | IV | V | VI | VII |
|---|---|---|---|---|---|---|---|
| GPU | 0.473 | 1.526 | 2.733 | 5.637 | 8.396 | 11.735 | 15.598 |
| CPU | 6.378 | 21.574 | 39.078 | 80.902 | 120.823 | 168.509 | 224.627 |

*4.4 Updating configurations*

Leap-Frog integrating consumes most time in updating the molecular configurations. Leap-Frog scheme can be parallelized as easy as the bonded interaction algorithms. Table X shows that the Leap-Frog GPU kernel also gets very good speedup as the bonded GPU kernels.

Table X. Updating configuration time (μs)

| System | I | II | III | IV | V | VI | VII |
|---|---|---|---|---|---|---|---|
| GPU | 0.101 | 0.268 | 0.464 | 0.888 | 1.34 | 1.834 | 2.425 |
| CPU | 1.993 | 6.771 | 12.637 | 27.623 | 41.467 | 57.670 | 76.868 |

To remove center-of-mass (COM) translation and rotation, parallel summation on GPU is needed. As shown in Table XI, the performance is not as good as the updating algorithm, which is ascribed to the two steps of summation and the parallelism of the second step is not high with only one thread in a block.

Table XI. Execution time (μs) of removing COM translation and rotation

| System | I | II | III | IV | V | VI | VII |
|---|---|---|---|---|---|---|---|
| GPU | 0.207 | 0.442 | 0.439 | 0.704 | 0.942 | 1.224 | 1.542 |
| CPU | 0.586 | 1.972 | 3.578 | 7.732 | 12.384 | 18.310 | 24.928 |

*4.5 Overall performance of the whole program*

After testing the performance of every step, the overall performance of the program is evaluated. The GPU/CPU speedup ratios according to numbers of atoms are depicted in Figure 6. The time consuming proportion of every step is different between GPU and CPU program. In CPU program, the computation of interactions takes most of the time, usually more than 70%. While in GPU program, neighbor list generation is the most time consuming procedure which is usually more than 50%. Although the speedup of neighbor list generation is only about 3 times of GPU over CPU, the speedup of the interactions procedure is very high. So the overall speedup of GPU program depicted in Figure 6 is higher than the GPU neighbor list generation procedure.

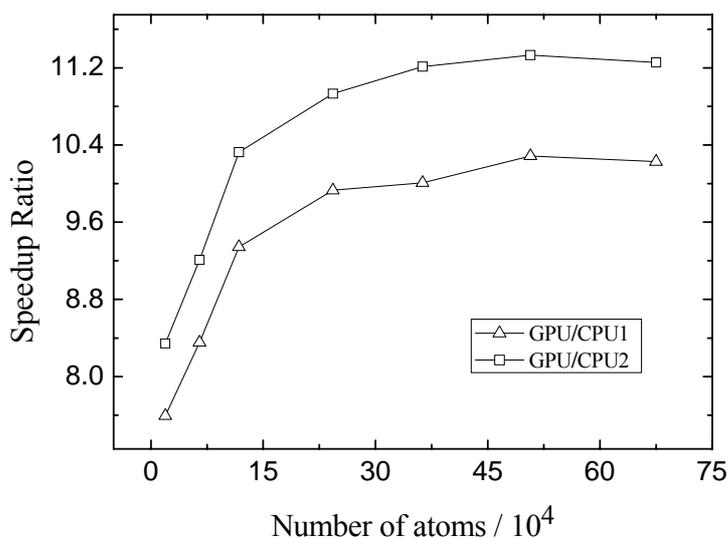

Figure 6. Over all speedup ratio of GPU/single CPU over system sizes

The GPU program is also compared against the GROMACS-4.0.5 parallel program running in one node with 8 processes. Domain decomposition can not be used with angular type of COM remove in this version of GROMACS, so particle decomposition is adopted instead. And the GPU speedup ratio against parallel CPUs is depicted in Figure 7. It can be seen that the speedup ratios are still going up with the system size.

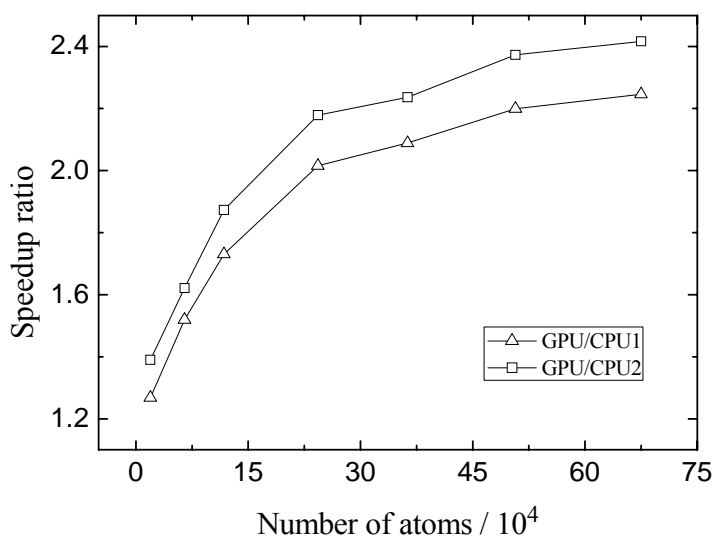

Figure 7. Overall speedup ratio of one GPU over 8 CPUs running GROMACS-4.0.5 under different system sizes

Although our GPU codes speed up about 10 times of single process GROMACS-4.0.5, there are still some aspects to be optimized. The first is the process of neighbor list generation, which takes about 20% of the total time when updated every 10 steps. Further speedup can be obtained if the charge group concept, as in GROMACS-4.0.5, can be introduced into our GPU codes. Secondly, in some algorithms of our GPU codes, the global memory reading is not in the coalesced way such as the exclusions data reading. In the inner loop of neighbor list generation, every thread must load the exclusions of the home atom. And some threads may read the same memory data of the GPU global memory. If the GPU shared memory is large enough, all the exclusions of dealing atoms in one thread block can be read into shared memory in advance. Third, branching operations should be minimized. In the calculations of angle and dihedral interactions, the interaction order must be defined according to the position of the angle or dihedral the atom locates, so several branching instructions have to be used in each thread, which affects the efficiencies of GPU kernels considerably. If writing into the same global memory of GPU without conflict can be realized in the future generation of GPU, and one angle or dihedral interaction will be managed by one thread, this problem will be solved.

**5 Analysis of simulation results and validation of the program**

The type of variables in GPU kernels is almost float because double precision operations execute much slower in current GPU [7]. And summing float type variables in a different order can lead to different results [9], which will affect the accuracy of force computation, kinetic energy summation, summation of translation and rotation COM. Accumulation of such errors after thousands or even millions of MD simulation steps may cause considerable difference in the trajectories between GPU and CPU programs. In order to validate our GPU computation, the simulation results from both programs for the same case are analyzed in detail.

Yu et.al.[23] have simulated the interpenetration and crystallization processes of a polyethylene (PE) system with 150 macromolecules composing 150 beads each using GROMACS 3.3[3] on clusters with Xeon and Itanium CPUs. In this paper, the system is enlarged by 8 times to 360,000 (1200 macromolecules composing 300 beads each) atoms. Two interpenetration cases and four crystallization cases are simulated and the detailed information is listed in Table XII.

Table XII. Simulation cases for interpenetration and crystallization.

| Case | Program | Atom number | Temperature (K) | Simulation time (ns) | ICF of initial structure |
|---|---|---|---|---|---|
| interpenetration | | | | | |
| IP_C | GROMACS-4.0.5 | 45,000 | 1000 | 30 | - |
| IP_G | GPU | 360,000 | | 30 | |
| crystallization | | | | | |
| C_1 | GROMACS-4.0.5 | 45,000 | 600 | 12 | Low |
| C_2 | | | | | High |
| G_1 | GPU | 360,000 | | 50 | Low |
| G_2 | | | | | High |

As our previous work[23], the interpenetration process of the PE chains is performed first. The auto-correlation function (ACF) of end-to-end vectors of IP_G system are calculated and compared with the IP_C system in ref. 21. The result is plotted in Figure 8. It can be found that in both IP_G and IP_C systems the ACF of end-to-end vectors declines to lower than 0.1 after about 5ns, indicating the same relaxation time in both systems. After being equilibrated, the conformation distribution of the IP_G system is compared to the IP_C system and shown in figure 9. One can find that they are almost overlapped, so that the bonded micro-structure simulated by GPU program is correct. Meanwhile, the normalized radial distribution function (RDF) of both IP_G and IP_C systems are compared (Figure 10) and it can be found that the appearance position of peaks on the RDF curve in the IP_G system is just the same as the IP_C system, indicating the correctness of non-bonded micro-structure obtained by GPU program. Further more, other properties of the interpenetration are calculated for both systems. Figure 11 shows the inter-chain contact fraction (ICF) during the interpenetration process in both systems. The ICF [23,24] is though to be the structural property of the entanglement in polymer system. This figure shows that both systems have almost the same ICF plateau. Figure 12 shows the average radius of gyration ($R_g$) of all chains during the interpenetration in both IP_G and IP_C systems. With confidence, the max $R_g$ in IP_G system is slightly larger than that in IP_C system, indicating that the final coil of IP_G system is even larger than IP_C system and less surface confinement is presented. All the comparison above shows that the GPU program can achieve the same result as the GROMACS/CPU package for the high temperature simulation.

When the temperature decreases, crystallization will take place in PE. The crystallization processes starts from initial structures with different ICF are also simulated in the present work as what have been performed previously[23]. Table XII shows the 2 different samples in the crystallization simulation by GPU. The ICF of initial structure is low in G_1 system while high in G_2 system. The simulation takes a much longer time (50 ns) than the C_1 (the L_3 in ref. 21) and C_2 (the H_1 in ref. 21) systems. Figure 13 shows the distribution of crystalline domains in the C_1 and G_1 systems by the SOP imagination method[23]. The green points in this figure represent the crystalline domains in the central slices at 2 ns of the crystallization processes. It can be seen from these images that the nuclei or the crystallization domains are similar in size in both systems. Further quantificational analysis is taken. The average stem length for C_1, C_2, G_1 and G_2 are plotted in figure 14. For the G_1 and G_2 systems, the final stem length increases to about 13 bonds and the systems with initial lower ICF structures (G_1) would perform a slightly faster increase in the stem length at first but finally go to the similar value with the other one. The final values for G_1 and G_2 are nearly the same as the C_1 and C_2, and the same trend that the system with lower ICF initial structure would perform faster increase first can be found in C_1 and C_2. Moreover, figure 15 shows the variation of fraction of trans-conformation during the crystallization processes in these four systems, and figure 16 shows the variation of crystallinity during the crystallization processes. In each figure, the same final values for G_1, G_2 as C_1, C_2 can be clearly found. And for the variation of the crystallinity the systems with lower initial structures (C_1, G_1) can perform higher value all the time than the systems with higher initial structures (C_2, G_2). While for the variation of the trans-conformation, the systems with lower initial (C_1, G_1) structures only perform higher value at first and later became the same as the systems with higher initial structures (C_2, G_2). For the tests above, the thermodynamics properties obtained by GPU program are the same as those obtained by GROMACS-4.0.5 simulation. Thus the GPU program is convinced to be valid for the simulation of crystallization.

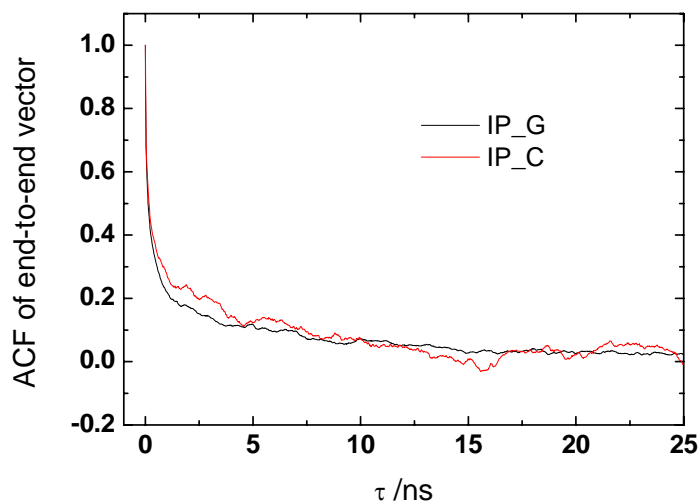

Figure 8. Auto-correlation function of end-to-end vectors for both IP_G and IP_C systems.

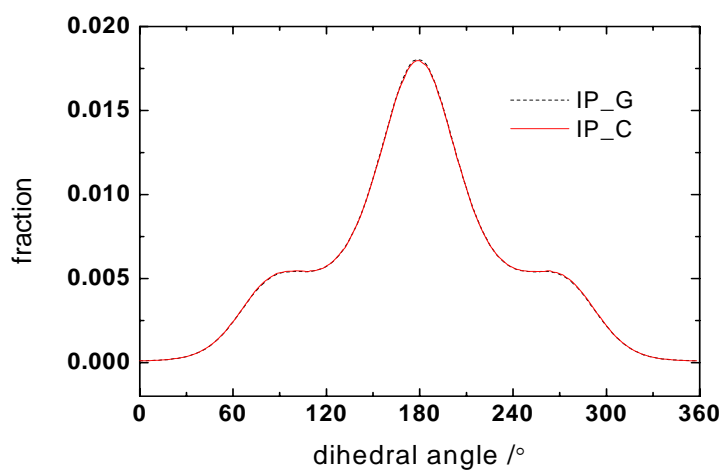

Figure 9. Conformation distribution after the interpenetration is equilibrated.

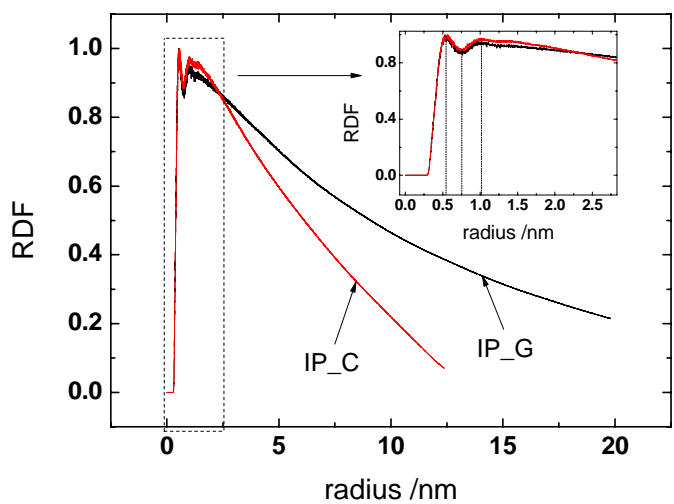

Figure 10. Radial distribution function (RDF) of the amorphous coils in both IP_C and IP_G systems.

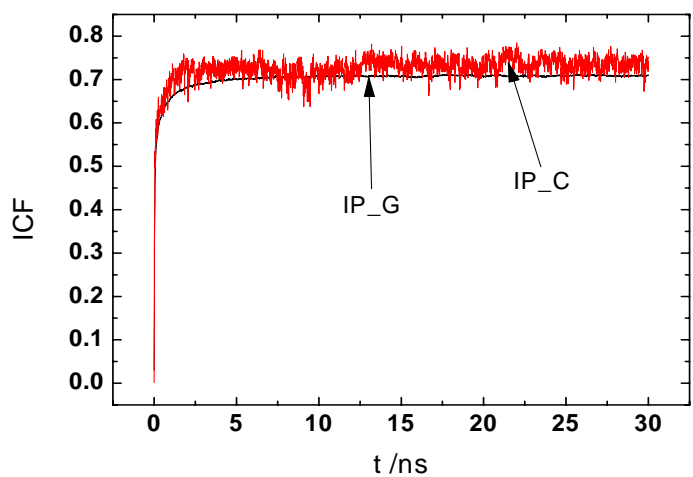

Figure 11. Inter-chain contact fraction (ICF) during the interpenetration process in both systems.

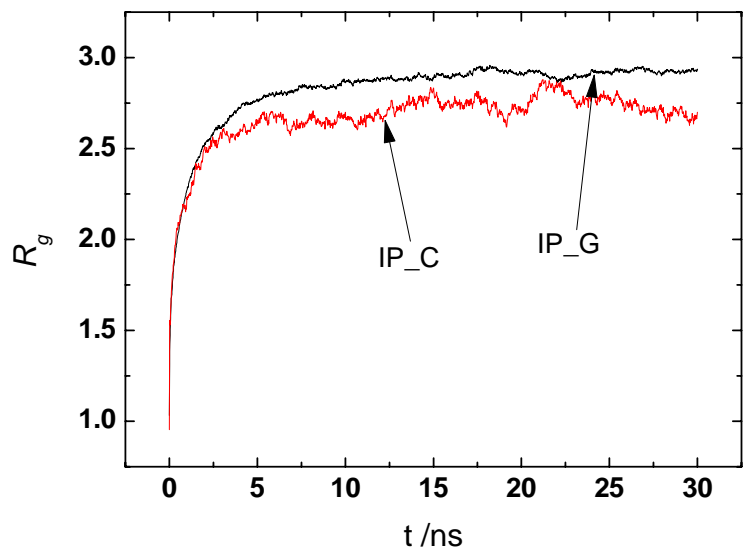

Figure 12. Variation of radius of gyration in both systems.

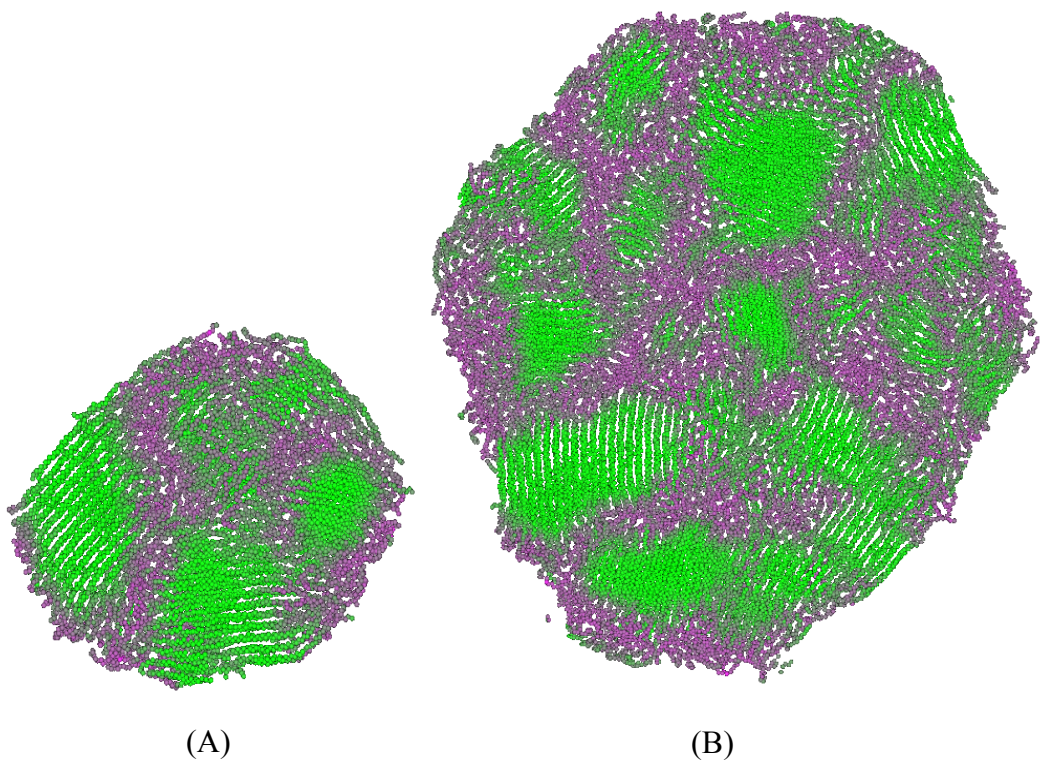

(A)  (B)

Figure 13. Distribution of crystalline domains and amorphous region by SOP images in the central slice at 2 ns of the crystallization processes in both systems. The green region represents the crystalline domains while the purple region represents the amorphous region. (A) sample C1; (B) the sample G1.

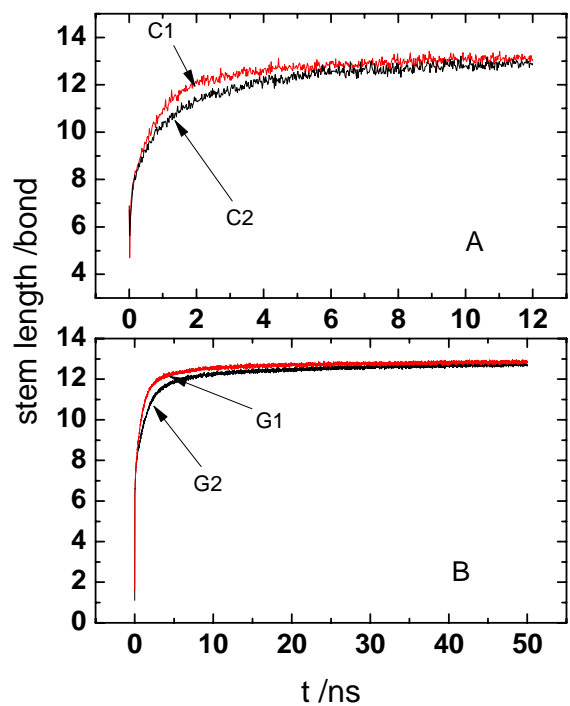

Figure 14. Average stem length during the crystallization processes.

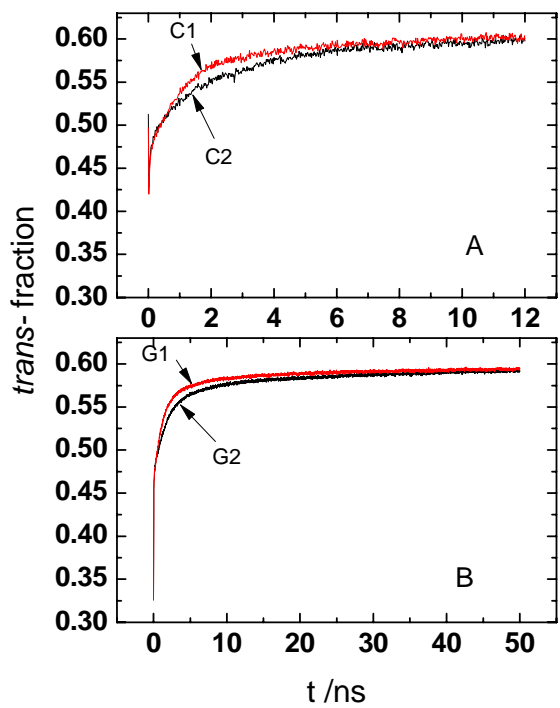

Figure 15. Fraction of trans- conformation during the crystallization processes.

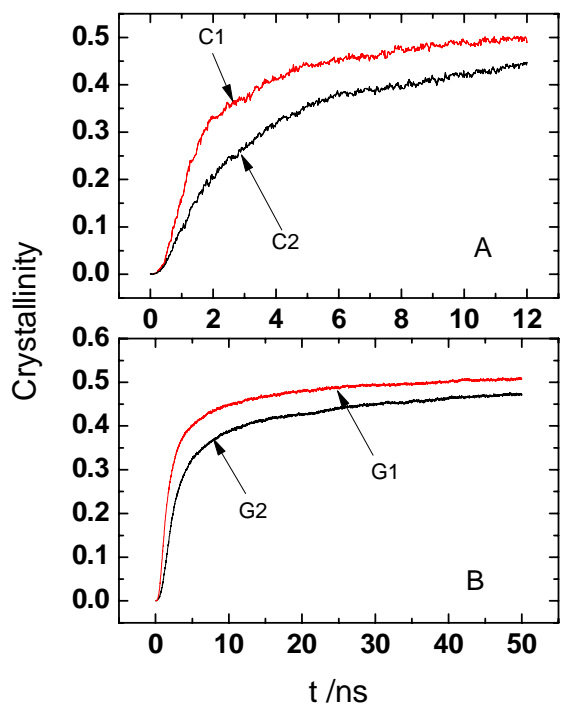

Figure 16. Variation of crystallinity with time.

**6 Conclusions and perspective**

In this paper, we presented macromolecular MD simulation codes fully implemented on GPU which reduces overheads of transferring data between GPU and CPU significantly. The performance of our GPU codes is compared against GROMACS-4.0.5. The GPU program on a NVIDIA GT200 GPU runs about 10 times faster than on a single core of Intel Xeon(R) E5430 2.66GHz CPU, and up to twice faster than 8 CPU cores in one node. We have also simulated a macromolecular system, which contains 1200 polyethylene molecules under NVT ensemble. The polymer crystallization phenomenon is observed and the studied physical features are reasonable. For polar linear polymer (where neighbouring particles along the polymer chain have alternating charges and are governed by long-range electrostatic interactions) systems[25], the Particle-Mesh-Ewald method of our previous work[26] can be integrated. However, for MD simulations of even larger systems, one GPU core is incompetent. Parallel GPU computations which uses the Message Passing Interface (MPI) library is under investigation.

**Acknowledgement**

This work is supported by the National Natural Science Foundation of China under the Grants Nos. 20874107 and 20221603 and Chinese Academy of Sciences under the Grant No. KJCX2-SW-L08.